\documentclass[epj]{svjour}

\usepackage{graphicx}   
\usepackage{dcolumn}    
\usepackage{bm}         

\begin{document}

\title{Optimal Investment Horizons}

\author{
        Ingve Simonsen\inst{1}\fnmsep\thanks{Email address: ingves@nordita.dk} \and
        Mogens H. Jensen\inst{2}\fnmsep\thanks{Email address: mhjensen@nbi.dk} \and 
        Anders Johansen\inst{2}\fnmsep\thanks{Email address: johansen@nbi.dk}} 

\institute{
  Nordic Institute for Theoretical Physics (NORDITA), Blegdamsvej 17, 
  DK-2100 Copenhagen {\O}, Denmark
\and 
 The Niels Bohr Institute, Blegdamsvej 17, DK-2100 Copenhagen {\O},
  Denmark
}

\date{\today}

\abstract{
  In stochastic finance, one traditionally considers the return as a
  competitive measure of an asset, {\it i.e.}, the profit generated by
  that asset after some fixed time span $\Delta t$, say one week or
  one year.  This measures how well (or how bad) the asset performs
  over that given period of time. It has been established that the
  distribution of returns exhibits ``fat tails'' indicating that large
  returns occur more frequently than what is expected from standard
  Gaussian stochastic processes~\cite{Mandelbrot-1967,Stanley1,Doyne}.
  Instead of estimating this ``fat tail'' distribution of returns,
  we propose here an alternative approach, which is outlined by
  addressing the following question: What is the smallest time
  interval needed for an asset to cross a fixed return level of say
  10\%?  For a particular asset, we refer to this time as the {\it
    investment horizon} and the corresponding distribution as the {\it
    investment horizon distribution}. This latter distribution
  complements that of returns and provides new and possibly crucial
  information for portfolio design and risk-management, as well as for
  pricing of more exotic options.  By considering historical financial
  data, exemplified by the Dow Jones Industrial Average, we obtain a
  novel set of probability distributions for the investment horizons
  which can be used to estimate the optimal investment horizon for a
  stock or a future contract.
}
\PACS{{}{}}
\mail{Ingve Simonsen}

\maketitle


Financial data have been recorded for a long time as they represent an
invaluable source of information for statistical investigations of
financial markets. In the early days of stochastic finance it was
argued that the distribution of returns (see definition below) of an
asset should follow a normal (Gaussian)
distribution~\cite{Book:Bouchaud-2000,Book:Mantegna-2000}.  However,
by analysing large, and often high-frequency, financial data sets, it
has been established that these distributions on short time scales ---
typically less then a month, or so --- can posses so-called
``fat-tails'', {\it i.e.}  distributions that show strong deviations
from that of a Gaussian~\cite{Mandelbrot-1967,Stanley1,Doyne} with higher
probabilities for large events.
This is similar to the distributions found for turbulence in air and
fluids which have led to comparisons between the statistics of
financial markets and that of turbulent
fluids~\cite{Book:Mantegna-2000,Stanley2,Peinke1,Peinke2}.  In
turbulence, one obtains stretched exponential distributions which find
their analogy in finance when considering higher order correlations of
the asset price~\cite{Anders1,Anders2}.

In order to get a deeper understanding of the fluctuation of financial
markets it is important to supplement this established information of
fluctuations in the returns with alternative measures.  In the present
paper, we therefore ask the following ``inverse'' question: ``What is
the typical time span needed to generate a fluctuation or a movement
(in the price) of a given
size''~\cite{MHJ,Luca,rome,Book:Karlin-1966,DR1,DR2}.  Given a fixed
log-return barrier, $\rho$, of a stock or an index as well as a fixed
investment date, the corresponding time span is estimated for which
the log-return of the stock or index {\it for the first time} reaches
the level $\rho$.  This can also be called the first passage time
through the level (or barrier)~\cite{Book:Karlin-1966,DR1,DR2,Maslov}
$\rho$.  As the investment date runs through the past (price) history
of the stock or index, the accumulated values of the first passage
times form the probability distribution function of the investment
horizons for the smallest time period needed in the past to produce a
log-return of at least magnitude $\rho$.  The maximum of this
distribution determines the most probable investment horizon which
therefore is the optimal investment horizon for that given stock or
index.

The first passage time is important from an economic point of view in
several ways.  Firstly, say an investor plans to sell or buy a certain
asset.  Then, of course, he or she is interested in doing the
transaction at a point in time that will optimize the potential
profit, {\it i.e.}  to sell for the highest possible price, or, for a
buyer, to buy for the lowest price.  However, the problem is that one
does not know when the price is optimal.  Therefore, the best one can
do, from a statistical point of view, is to make a transaction at a
time that is probabilisticly favorable.  This optimal time, as we will
see, is determined by the maximum of the first passage time
distribution, {\it i.e.} {\em the most likely first passage time}.
Secondly, for a holder of an European type option, either a call or a
put of given strike price, the most likely first passage time will, in
much the same way as presented above, define the optimal maturity of
the option. Furthermore, for an American type call option the most
likely first passage time of the underlying asset will be useful to
know when to exercise the option.  These same arguments apply even
more to exotic options used in the financial industry~\cite{Exotic}.
Thirdly, the investment distribution for {\em negative} levels of
returns, provides crucial information for the implementation of
certain {\em stop-loss} strategies.  Finally, but not least, the first
passage distribution will by itself give invaluable, non-trivial
information about the stochasticity of the underlying asset price.
This point has recently been demonstrated explicitly in the (related)
field of turbulence~\cite{MHJ,Luca}.

\begin{figure}[tbp]
  \begin{center}
    \leavevmode 
    \includegraphics[width=0.75\columnwidth,angle=-90]{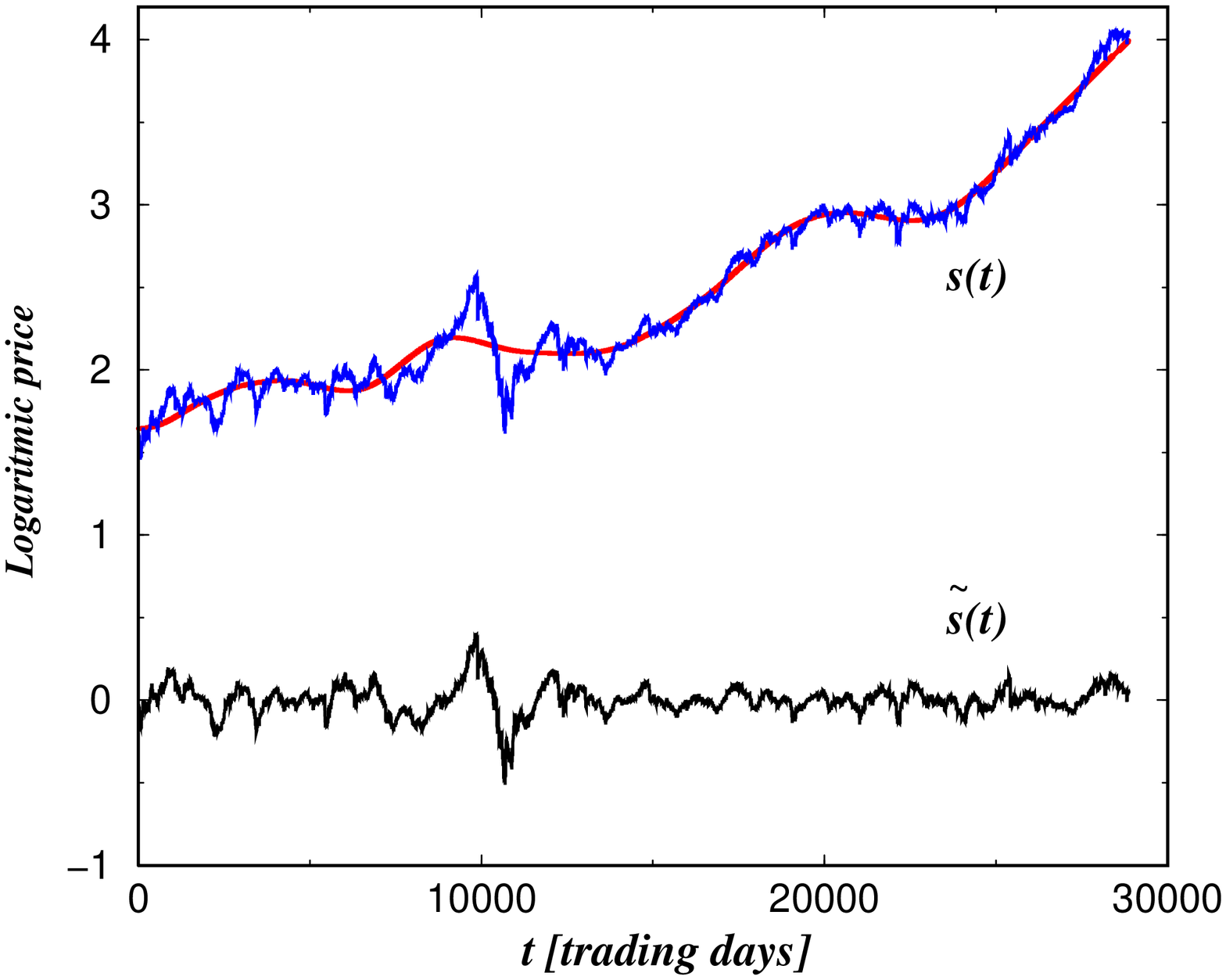} 
    \caption{
      The historic daily logarithmic closure prices of the Dow Jones
      Industrial Average~(DJIA) over the period from May 26, 1896 to
      June 5, 2001. The upper curly curve is the raw logarithmic DJIA
      price $s(t)=\ln S(t)$, where $S(t)$ is the historic daily
      closure prices of the DJIA. The smooth curve represents the
      drift on a scale larger then $1000$ trading days. This drift was
      obtained by a wavelet technique as described in the main text.
      The lower curly curve represents the wavelet filtered
      logarithmic DJIA data, $\tilde{s}(t)$. Those filtered data are
      just the fluctuation of $s(t)$ around the drift.}
    \label{fig:djia}
  \end{center}
\end{figure}

To illustrate our procedure we consider the Dow Jones Industrial
Average (DJIA).  We analyze the daily closure of the DJIA from its
beginning on May 26, 1896 to June 5, 2001~(present). This leave us
with $105$ years of data, or almost $30\,000$ trading days. The data
set to be analysed is depicted in Fig.~\ref{fig:djia}.  The log-return
over a time interval $\Delta t$ of an asset of price $S(t)$, at time
$t$, is defined as
\begin{eqnarray}
    \label{Return}
    r_{\Delta t}(t) &=& s(t+\Delta t)- s(t),
\end{eqnarray}
where $s(t) = \ln S(t)$, {\it i.e.}, the log-return is just the
log-price change of the asset.  The investment horizon,
$\tau_\rho(t)$, at time $t$, for a return level $\rho$ is defined at
the smallest time interval, $\Delta t$, that satisfies the relation
$r_{\Delta t}(t)\geq \rho$, or in mathematical
terms~\cite{Book:Karlin-1966} $\tau_\rho(t) = \inf \left\{ \Delta
  t>0 \,| \,r_{\Delta t}(t)\geq \rho \right\}$.  The investment
horizon distribution, $p(\tau_\rho)$, is obtained as the histogram of
investment horizons $\tau_\rho$.  Furthermore, we introduce the
cumulative distribution for the horizon being larger then $\tau_\rho$,
{\it i.e.}
\begin{eqnarray}
    \label{Cumulative}
    P(\tau_\rho) &=& \int_{\tau_\rho}^\infty p(t)\, dt.
\end{eqnarray}

It is well-known, and easy to see from Fig~\ref{fig:djia}, that there
is a substantial upward drift with time in the DJIA which is an
indication of the overall growth of the world economy.  In order to
reduce this drift, we have wavelet filtered~\cite{Book:NR-1992} the
data on a scale larger then $1000$ trading days, and we will therefore
in this study limit ourselves to the behavior up-to $1000$ trading
days (about $4$ years). This is achieved by first transforming the
log-price $s(t)=\ln S(t)$ to the wavelet-domain, setting all wavelet
coefficient corresponding to scales larger then $1000$ trading days to
zero, and finally transforming back to the time domain. This
procedure, which results in the filtered log-price time series
$\tilde{s}(t)$, will reduce the effect of drift as can be seen
explicitly from Fig.~\ref{fig:djia}.  
In the following analysis we
will therefore use $s(t)$ to denote the {\em filtered} time series,
and subsequently $S(t) = \exp(s(t))$ for the filtered price.

\begin{figure}[tbp]
  \begin{center}
    \leavevmode 
    \includegraphics[width=0.75\columnwidth,angle=-90]{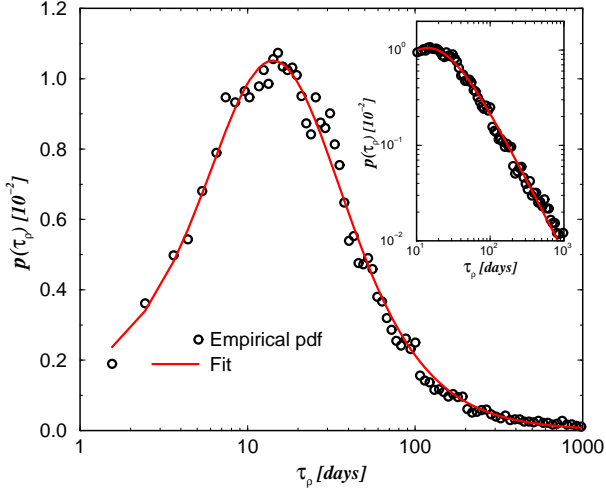} 
    \caption{
      The probability distribution function~(pdf), $p(\tau_\rho)$, of
      the investment horizons (first passage times) measured in
      trading days, $\tau_\rho$, at a level $\rho=0.05$ ({\it i.e.}
      $5\%$ return).  The data used to produce this figure are the
      wavelet filtered logarithmic returns calculated from the
      historic daily closure prices~(Fig.~\protect\ref{fig:djia}).
      The open circles represents the empirical pdf (at level $\rho$).
      We notice the pronounced maximum of this function at
      approximately $\tau_\rho^*=15$~trading days (note the log-linear
      scale).  This maximum represents the most probable time of
      producing a return of $5\%$.  The solid line represents a
      maximum likelihood fit to the functional
      form~\protect(\ref{fit-func}) with parameters: $\alpha=0.50$,
      $\beta=4.5$~days$^{1/2}$, $\nu=2.4$, and $t_0=11.2$~days. The
      inset is the same figure on log-log scale such that the
      power-law behavior of the tail is more easily observed.}
    \label{fig:pdf-0.05}
  \end{center}
\end{figure}

To present our analysis we show in Fig.~\ref{fig:pdf-0.05} the
investment horizon distribution $p(\tau_\rho)$ for a reasonably large
value of the return level, $\rho=0.05$ ({\it i.e.} $5\%$). This figure
indicates that the obtained distribution has a very interesting and
unexpected form. It exhibits a rather well defined and pronounced
maximum, followed by an extended tail for very long time horizons
indicating a non-zero and important probability of large passage times
(note that the $\tau_\rho$-axis is logarithmic).  We expect that these
long investment horizons reflect periods where the market is
reasonably calm and quiet, or is going down for a long period of time
before finally coming back up again. The short horizons on the other
hand -- those around the maximum -- are observed in more volatile
periods, which appears to be the most common scenario. Indeed the
maximum obtained at $\tau_\rho =\tau_\rho^{*}$ is the most likely
horizon, which we call the {\em optimal investment horizon}. This kind
of distribution of the investment horizon statistics for economical
markets has, to the best of our knowledge, not been published before.

\begin{figure}[tbp]
  \begin{center}
    \leavevmode
    \includegraphics[width=0.75\columnwidth,angle=-90]{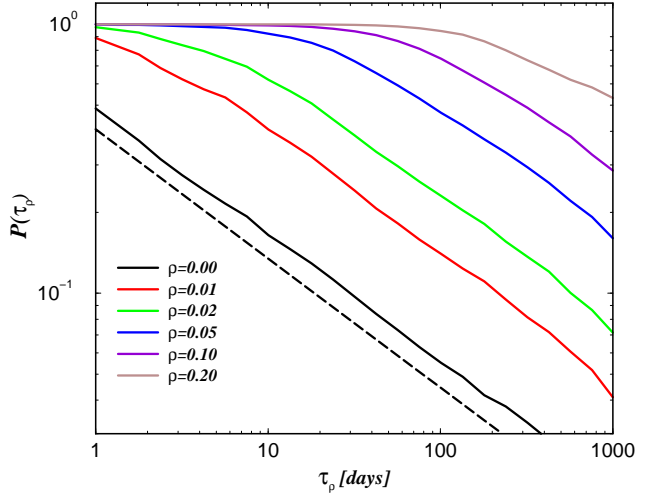}
    \caption{
      The empirical cumulative probability distributions (solid
      lines), $P(\tau_\rho)$, vs. horizon $\tau_\rho$ for different
      levels $\rho=0$, $0.01$, $0.02$, $0.05$, $0.10$, and $0.20$
      (from bottom to top), for the (wavelet filtered) Dow Jones
      Industrial Average.  The dashed line corresponds to the
      geometrical Brownian motion assumption for the underlying price
      process, in which case one should have
      $P(\tau)\sim\tau^{-\alpha}$ with a tail index $\alpha=1/2$. For
      the level $\rho=0$ the geometrical Brownian motion assumption
      seems reasonable ({\it i.e} $\alpha_o \simeq 1/2$), while the
      tail index, $\alpha_\rho$, shows only a weak (if any) dependence
      on the level $\rho$. }
    \label{fig:cdf-log}
  \end{center}
\end{figure}

To better understand the tail of this distribution, we consider a
rather small level of return $\rho$. If this level is small enough, it
is likely that the return will break through the level after the first
day, while larger horizons will become more an more unlikely.
However, the probability for a large horizon will not be zero; if,
say, we consider a positive (but still small) level $\rho$, then a
period of recession will result in a $\tau_\rho(t)$ that might be
considerably larger then one day since it takes time to recover after
a set back. For instance, after the 1927 stock market crash, it took
more then two decades for the DJIA to regain the value it had just
before the crash. In the limit $\rho\rightarrow 0^+$ the horizon
distribution $p(\tau_0)$ is known in the literature as the {\em first
  return} probability distribution for the underlying stochastic
process~\cite{Book:Karlin-1966,DR1}. In Fig.~\ref{fig:cdf-log} (lowest
solid curve) the cumulative distribution, $P(\tau_0)$, for the DJIA
data set is shown.  It is observed that the tail of this distribution
scales as a power law, $P(\tau_0)\sim \tau_0^{-\alpha_0}$, with
$\alpha_0\simeq 1/2$, over almost three orders of magnitudes in time.
This value of the exponent can be understood as follows: If we assume,
as is common in the financial
literature~\cite{Book:Bouchaud-2000,Book:Mantegna-2000}, that the
asset price, $S(t)$, can be well approximated by a geometrical
Brownian motion, then trivially it follows that $s(t)=\ln S(t)$ will
just be an ordinary Brownian motion.  From the literature it is
well-known that the first-return probability distribution of a
fractional Brownian motion scales like~\cite{Book:Karlin-1966,DR1}
$p(\tau)\sim \tau^{H-2}$, where $H$ is the Hurst exponent, and hence
$P(\tau)\sim \tau^{H-1}$.  Since the Hurst exponent of an ordinary
Brownian motion is $H=1/2$, the empirically observed scaling (see
Fig.~\ref{fig:cdf-log}) $P(\tau_0)\sim \tau_0^{-1/2}$ is a consequence
of the (at least close to) geometrical Brownian motion behavior of
$S(t)$. This argument of an unbiased geometrical Brownian motion is
also strengthen by observing that $P(\tau_0=1)\simeq 1/2$, meaning
that the log-price change over one day raises half of the time and
drops in the remaining half.  It should noticed that in order to
observe the power-law of exponent $\alpha_0=1/2$ for small levels, the
filtered data have to be used. Using the raw data would, due to the
presence of the drift, result in a slightly larger (smaller) exponent
for a positive (negative) small level $\rho$.

Fig.~\ref{fig:cdf-log} also shows the cumulative distributions
$P(\tau_{\rho})$ for different choices of the return level $\rho$,
{\it i.e.},  $\rho = 0.01$, $0.02$, $0.05$, $0.10$ and $0.20$. From
this figure it is seen that the tail exponent, $\alpha_\rho$, is
rather insensitive to the return level. In particular one finds that
$\alpha_\rho\simeq 1/2$ over a broad range of values for $\rho$, a
value that is consistent with the geometrical Brownian motion
hypothesis of the underlying asset price process~(see
Eq.~(\ref{fist-passage-pdf}) below). Moreover, it is observed that as
the level $\rho$ is increased from zero, the most likely horizon moves
away form $\tau_0=1$ and toward larger values.  In other words ---
there is an {\em optimal investment horizon}, $\tau_\rho^{*}$,
corresponding to, and depending on, a given level of return $\rho$.
Furthermore, we have checked, and found, that there is an approximate
symmetry under $\rho\rightarrow -\rho$ for the investment distribution
as long as the filtered data are used. One therefore does not have to
consider negative levels explicitly. On the other hand, if the
analysis is based on the raw data there is a clear asymmetry.

The time needed for a general time series to reach a certain level is
in the mathematical literature known as the {\em first passage
  problem}.  For a Brownian motion this problem has been solved
analytically \cite{DR1,DR2}
\begin{eqnarray}
  \label{fist-passage-pdf}
    p(t) &=& \frac{1}{\sqrt{\pi}}\, \frac{a}{t^{3/2}} 
         \exp\left( -\frac{a^2}{t} \right). 
\end{eqnarray}
%
%
where $a$ is proportional to the level $\rho$.  For large times one
recovers, from Eq.~(\ref{fist-passage-pdf}), the well known
distribution of first return times to the origin $p(t) \sim t^{-3/2}$.
In order to fit a functional form to the distribution in
Fig.~\ref{fig:pdf-0.05} we generalize this expression and suggest the
following form
\begin{eqnarray}
    \label{fit-func}
    p(t) &=&
    \frac{\nu}{\Gamma\left(\frac{\alpha}{\nu}\right)}\,
    \frac{\beta^{2\alpha}}{(t+t_0)^{\alpha+1} }
    \exp\left\{
          -\left(\frac{\beta^2}{t+t_0}\right)^{\nu} 
        \,\right\},
\end{eqnarray} 
%
%
which reduces to Eq.~(\ref{fist-passage-pdf}) in the limit when
$\alpha=1/2$, $\beta=a$, $\nu=1$, and $t_0=0$, since
$\Gamma(1/2)=\sqrt{\pi}$.  The form~(\ref{fit-func}) seems to be a
good approximation to the empirical investment horizon distributions.
The shift $t_0$ is needed in order to fit the optimal horizon well,
and its actual value may depend on possible short-scale drift.  The
full-drawn curve in Fig.~\ref{fig:pdf-0.05} shows a (maximum
likelihood) fit to the empirical data with the functional form
Eq.~(\ref{fit-func}), and the agreement is observed to be excellent
(the fitted parameter values are given in the caption of
Fig.~\ref{fig:pdf-0.05}).
 
\begin{figure}[tbp]
  \begin{center}
    \leavevmode
    \includegraphics[width=0.75\columnwidth,angle=-90]{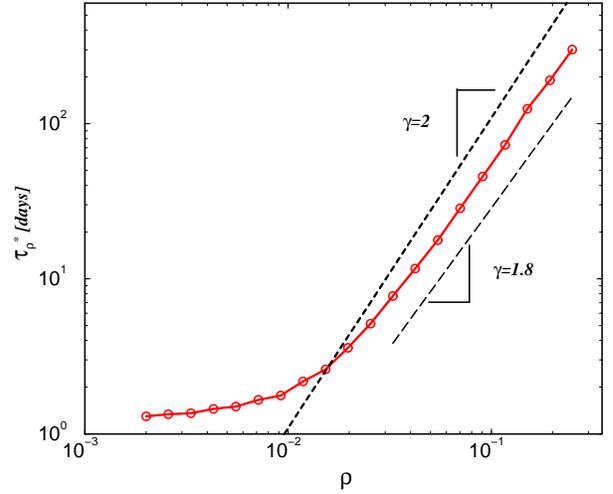}
    \caption{
      The optimal investment horizon, $\tau_\rho^*$, as a function of
      the log-return level $\rho$. The open circles represent the
      empirical results obtained from the (wavelet filtered)
      historical Dow Jones data.  The dashed line, of slope
      $\gamma=2$, corresponds to the geometrical Brownian motion
      hypothesis for the underlying asset price. One observes that for
      small levels this hypothesis fails dramatically. However, for
      levels of the order of a few percents or larger, the geometrical
      Brownian motion assumption becomes more realistic, {\em but}
      also in this region a discrepancy is observed.  Empirically one
      finds a scaling $\tau_\rho^*\sim \rho^\gamma$ with $\gamma\simeq
      1.8$~(long dashed line).}
    \label{fig:fit}
  \end{center}
\end{figure}

As the optimal horizon provides an important information for an
investor, a relevant question would now be: How does the optimal
horizon, $\tau_\rho^{*}$, depend on the return level $\rho$ ? This
dependence, as measured from the empirical horizon distribution, is
shown in Fig.~\ref{fig:fit}. Intuitively, it is clear that the optimal
horizon will increase rather rapidly as the return is increased.
However, we observe that this increase occurs in a systematic fashion
\begin{eqnarray}
    \label{optimal}
    \tau^*_\rho &\sim& \rho^\gamma,
\end{eqnarray}
with $\gamma\simeq 1.8$, see Fig.~\ref{fig:fit}. For a Brownian
motion, with a first passage distribution
Eq.~(\ref{fist-passage-pdf}), it is straightforward to derive that the
power law exponent should be $\gamma = 2$ for the {\em whole} range of
$\rho$.  Not surprisingly, we therefore find a deviation from standard
theories for the variation of the optimal horizon with the return
level. Furthermore, we systematically find $\nu>1$ when $\rho\neq 0$,
which is also inconsistent with the geometrical Brownian hypothesis.

In conclusion, we have obtained a new set of distributions of
investment horizons for returns of a prescribed level.  We have found
that the (empirical) optimal investment horizon depends on the return
level in a systematic fashion, which is not consistent with the the
geometrical Brownian motion hypothesis typically assumed in
theoretical finance~\cite{Book:Bouchaud-2000,Book:Mantegna-2000}. The
obtained distributions as well as the variation of the optimal horizon
can be applied if one wants to estimate the most probable time period
needed to stay in the market if an investor aims at a specific optimal
return.  Similar passage time distributions are found in turbulence of
fluids (where they are called inverse structure functions).  It
indicates that these distribution functions of passage times could be
a general and important feature of systems which exhibit extreme
events like for example in finance, turbulence, earthquakes,
and avalanches in granular media.

\section*{Acknowledgment}
  We thank E.~Aurell and S.~Maslov for valuable comments and suggestions.



\clearpage



\end{document}